\documentclass[preprint2]{aastex}

\usepackage{graphicx} 
 
\usepackage{natbib} 
 
 \usepackage{boxedminipage}

 \providecommand{\Irr}[1]{\ensuremath{\Phi^{#1}}} 
 \providecommand{\SEM}[1][]{\ensuremath{\mathrm{SEM}_{#1}}} 
 
\shorttitle{Helium line formation and abundance} 
\shortauthors{P. Mauas et al.}

\begin{document} 
\title{Helium line formation and abundance in a solar active region}  
 
\author{P.J.D. Mauas}  
 \affil{Instituto de Astronom\'\i a y F\'\i sica del Espacio,  
        Buenos Aires, Argentina } 
 \email{pablo@iafe.uba.ar} 
\author{V. Andretta}  
 \affil{INAF-Osservatorio Astronomico di Capodimonte,  
        Salita Moiariello 16, 80131 Napoli, Italy} 
\author{A. Falchi}  
 \affil{INAF-Osservatorio Astrofisico di Arcetri,  
        Largo Enrico Fermi 5, 50125 Firenze, Italy} 
\author{R. Falciani}  
 \affil{Dipartimento di Astronomia e Scienza dello Spazio, 
        Universit\`{a} di Firenze,  
        Largo Fermi 5, 50125 Firenze, Italy} 
\author{L. Teriaca} 
\affil{Max-Planck-Institut f\"{u}r Sonnensystemforschung 
Max-Planck-Str. 2, 37191 Katlenburg-Lindau, Germany} 
\and  
\author{G. Cauzzi} 
 \affil{INAF-Osservatorio Astrofisico di Arcetri,  
        Largo Enrico Fermi 5, 50125 Firenze, Italy}

\begin{abstract} 
An observing campaign (SOHO JOP 139), coordinated between ground based and SOHO 
instruments, has been planned to obtain simultaneous spectroheliograms of the 
same active region in several spectral lines. The chromospheric lines \ion{Ca}{2}~K, 
H$\alpha$ and \ion{Na}{1}~D as well as \ion{He}{1} 10830, 5876, 584 and 
\ion{He}{2}~304 \AA~lines have been observed. The EUV radiation in the 
range $\lambda<500$~\AA\ and in the range 
$260<\lambda<340$~\AA\  has  also been measured at the same time.  
These simultaneous observations allow us to 
build semi-empirical models of the chromosphere and low transition region 
of an active region, taking into account the 
estimated total number of photoionizing photons impinging on the 
considered active region and their spectral distribution. 
We obtained a model that matches very well all the observed line 
profiles, using a standard value for the He abundance ([He]=0.1) and a modified 
distribution of microturbulence. 
For this model we study the influence of the coronal radiation on the 
computed helium lines. We find that, even in an active region, the 
incident coronal radiation has a limited effect on the UV He lines, 
while it results of fundamental importance for the D$_3$ and 10830 lines.  
Finally we build two more models assuming values of He abundance [He]= 
0.07 and 1.5, only in the region 
where temperatures are larger than $1.\times 10^4$ K. This region, 
between the chromosphere and transition region, has been indicated as a good 
candidate for processes that might be responsible for strong variations of 
[He] \citep{Lie-Svendsen-etal:03}.  
The set of our observables can still be well reproduced in both 
cases changing the atmospheric structure mainly in the low transition 
region. This implies that, to choose between different values of  
[He], it is necessary to constrain the transition region with  
different observables, independent on the He lines. 
\end{abstract} 
 
\keywords{Sun:activity --- Sun:abundances --- Sun:Helium} 
 
\section{Introduction}\label{intro}

Helium is a very peculiar element in the sense that, despite its high 
abundance in the universe, its photospheric solar abundance is 
unknown.  Unfortunately it is undetectable in the visible photospheric 
spectrum and, until recently, the so-called solar photospheric 
abundance has been derived using theoretical stellar evolution models. 
A commonly accepted value for the ratio [He] = N$_{He}$/N$_{H}$ is 0.1 
and represents the helium abundance of the nebula from which the solar 
system formed.  However, there are now strong indications that this 
value is too large:  the inversion of helioseismic data by different 
authors leads to values of [He], in the convection zone, in the range 
0.078 -- 0.088 \citep[for a review, see][]{Boothroyd-Sackmann:03}.  A 
reduction of He abundance of roughly 10\% below its initial value could 
be explained including diffusion in the theoretical evolution models 
\citep{Proffitt-Michaud:91,Boothroyd-Sackmann:03}. 
 
The values quoted above refer to the outer convection zone and to the 
photosphere.  Measurements of [He] in the solar wind and interplanetary 
medium exhibit a high variability, from almost negligible values 
($\approx 0.01$) in equatorial regions, to high values ($\approx 0.3$) 
in transient events.  An average value in the fast component of the 
solar wind is about 0.05, thus about a factor of two lower than the 
photospheric [He] \citep{Barraclough-etal:95,Bochsler:98}.  This value 
is also in contrast with the general pattern of abundance variations in 
the fast wind, where elements with First Ionization Potential (FIP) 
$\geq 10$~eV typically retain their photospheric abundance, as opposed 
to low FIP elements ($\leq 10$~eV) which appear enhanced \citep[for a 
review, see][]{Meyer:96}. 
 
Direct measurements of [He] in the solar atmosphere closer to the surface 
are available only 
for the corona. 
In the quiet corona, at $R \approx 1.1\: R_{\sun}$,  
\cite{Gabriel-etal:95} and \cite{Feldman:98} found  
[He]~$\approx 0.08$, 
suggesting no helium depletion in the low corona, while in 
a equatorial streamer, at $R \approx 1.5\: R_{\sun}$,  
\cite{Raymond-etal:97}  
found an upper limit of 0.05.  
More recently,  
\cite{Laming-Feldman:01,Laming-Feldman:03} 
have estimated an abundance   
of 0.05 and 0.04 respectively, at distances $R<1.1\: R_{\sun}$. 
 
Apart from the observed variability and/or inconsistencies of the  
measurements in the corona and the solar wind, the general pattern  
of the FIP effect suggests that elemental fractionation should  
occur in the chromosphere, at $T<10^4$~K  
\citep{Geiss:82}.   
Detailed theoretical calculations including diffusion effects  
\citep{Hansteen-etal:97,Lie-Svendsen-etal:03} seem to confirm that,  
indeed, [He] 
undergoes strong changes from the chromosphere to the  
lower corona, before stabilizing at the values observed in  
the solar wind. 
 
Therefore it would be very important to determine the 
He abundance at chromospheric and transition region (TR) levels 
directly from line profiles. 
The strongest He lines in the solar atmosphere are in the extreme 
ultraviolet (EUV) below 584 \AA. 
From the ground, the only observable He line in the quiet solar atmosphere 
(on the disc) 
is in the near infrared at 10830 \AA, but when active regions or flares are 
observed the He lines are enhanced and in particular the $D_3$ line at  
5876 \AA~can also be detected. The excitation of these subordinate lines is 
at least partially due to a photoionization-recombination mechanism or ``PR'' 
process \citep{Zirin:75,Andretta-Jones:97}, 
i.e. to 
direct photoionization of chromospheric helium atoms by coronal EUV  
radiation shortward of 504 \AA\ 
and successive recombination to the excited levels of \ion{He}{1}. 
Observations actually suggest that this  process might not be 
very important in quiescent regions, at least for the 
strong resonance, hydrogen-like \ion{He}{2} lines such as the Lyman-$\alpha$  
at 304~\AA~\citep[e.g.:][]{FAL:93,Andretta-etal:00,Andretta-etal:03}.   
However,  no conclusive 
determination of the role of this process is available yet for active 
regions, where the coronal EUV radiation field is much more intense than in 
quiet areas.   
Hence, the study of the solar He spectrum requires not only spectral 
observations of an active region (or flare) in a very large spectral range  
(from EUV to near infrared), but also an estimate of the coronal EUV radiation 
impinging on the observed active region.  
 
To this aim  
we planned an observing campaign (SOHO JOP 139) coordinated between 
ground based and SOHO instruments  
to obtain simultaneous spectroheliograms of the same area in 
several spectral lines, including four He lines, that sample the solar  
atmosphere from the chromosphere to the transition region. During this 
campaign we observed the region NOAA 9468 (cos$\theta$=0.99) on May 26,  
2001 from 13:00 to 18:00 UT.  A small two-ribbon flare (GOES class C-1), 
developed in this region around 16 UT and its dynamics during the 
impulsive phase has been already studied in  
\cite{Teriaca-etal:03}. 
 
In this paper we concentrate on the pre-flare phase. We present  
semi-empirical models of the atmosphere of the active region (at 15:20 UT) 
constructed to match 
the observed line profiles from the chromosphere to the TR 
and taking into account the EUV radiation in the range 1 -- 500 \AA.  
The comparison between the observed and computed He line 
profiles allowed us to test the relevance of the PR process and of the  
He abundance values in an active region. 
In a following paper we will study the formation of He lines in the flare.

\section{Observations}\label{obs} 

A detailed description of the observing program is given in  
\cite{Teriaca-etal:03}.  We recall here that spectroheliograms were 
acquired with the Horizontal Spectrograph at the Dunn Solar Telescope (DST) of 
the National Solar Observatory / Sa\-cra\-men\-to Peak in the chromospheric 
lines Ca~{\sc ii} K, H$\alpha$ and Na~{\sc i} D as well as in the 
\ion{He}{1} lines at 5876 (D$_3$) and 10830 \AA.  The full field of view of the 
DST, 170$\arcsec\times 170\arcsec$, was covered in about 5 minutes, with a 
sampling step of $2\arcsec$. Correcting for offsets among 
the different detectors resulted in a final useful field of view (FOV) of 
$160\arcsec\times 140\arcsec$ with an effective resolution of $2\arcsec$. 
 
During the same period, spectroheliograms of the active region were obtained in 
the spectral windows around the \ion{He}{1} 584 \AA\ and the \ion{He}{2} 304 \AA\ 
spectral lines with the Normal Incidence Spectrometer (NIS) 
of the Coronal Diagnostic Spectrometer (CDS) aboard 
SOHO. The slit was stepped $6\arcsec$  
covering a $148\arcsec$ wide area in $\sim 5.5$ minutes. The final 
useful FOV was $148\arcsec\times 138\arcsec$ with an effective resolution of 
$6\arcsec\times 3.4\arcsec$. 
Ground-based and CDS data were aligned using SOHO/MDI images as a reference. 
We estimate the error around few arcseconds. 
CDS and ground based spectra are  
simultaneous within 2 minutes.

The EUV flux 
was monitored by the Solar EUV Monitor (\SEM) instrument 
aboard SoHO \citep{Hovestadt-etal:95}. 
The CELIAS/\SEM\ instrument  
provides calibrated total photon counts in the range $\lambda<500$~\AA\ 
(zero-th order, \SEM[0]), and in the range $260<\lambda<340$~\AA\ (first order, 
\SEM[1]), at 1~AU.   
 
Finally, the Extreme ultraviolet Imaging Telescope  
\citep[EIT;][]{Delaboudiniere-etal:95} 
provided  synoptic series of full disk images 
centered on  171, 195, 284 and 304 \AA\ around 13:00 and 19:00 UT.

 
\section{Derivation of the photoionizing radiation}\label{calc:photorad} 
 
In Sec.~\ref{intro} we have mentioned the role of EUV radiation, 
emitted mainly by coronal plasma, in the excitation of helium lines.   
The relevant radiation we need to estimate falls in the wavelength range below 
the photoionization threshold of \ion{He}{1} at 504~\AA\ (the photoionization 
threshold for \ion{He}{2} is at 228~\AA).  A misconception, commonly found in 
the literature, is that it would be sufficient to have an estimate of the 
intensities just near the photoionization thresholds, because of the sharp 
decrease with wavelength of the photoionization cross-sections -- which are 
proportional to $\approx \lambda^2$ and to $\approx \lambda^3$ for \ion{He}{1} 
and \ion{He}{2}, respectively.  However, as discussed more in detail by 
\cite{Andretta-etal:03}, even photons of considerably shorter wavelength can 
efficiently photoionize helium atoms and ions: what matters most is the 
\emph{ratio} between helium and hydrogen photoionization cross sections, at 
least for $\lambda>20$--$50$~\AA\ (at shorter wavelengths, inner-shell 
photoionization of metals absorb most of the photons).  The wavelength of the 
photoionizing photons mainly determines how deeply in the atmosphere they can 
penetrate, since the photoionization absorption cross-sections decrease with wavelength. 
This point will be further illustrated in Sec.~\ref{calc:atmos:EUVspectrum}. 
 
Therefore, for the calculation of the He line profiles we needed to estimate 
the total number of photoionizing photons impinging on the target active 
region, plus their spectral distribution. 
We obtained reliable estimates of these quantities combining both 
observational constraints from CELIAS/\SEM\ and EIT, and theoretical  
tools. The adopted procedures are illustrated in the following 
sections. 
 
\subsection{Coronal irradiance}%
 \label{calc:photorad:irradiance} 
 
The solar irradiances measured at about 15:20~UT  
in the \SEM[0] and \SEM[1] wavebands are 
$\Irr{\SEM[0]}=5.05\times 10^{10}$ 
photons cm$^{-2}$ s$^{-1}$ and $\Irr{\SEM[1]}=2.57\times 10^{10}$ photons 
cm$^{-2}$ s$^{-1}$, respectively, %
  with noise-like variations of the order of 0.1\%. Even the subsequent 
  C1.1-class flare, starting at 16:01 UT \citep{Teriaca-etal:03}, 
  produced variations in the \SEM\ irradiances limited to at most 1\%.   
  Both quantities include the contribution of the \ion{He}{2}~304~\AA\ 
  line, while $\Irr{\SEM[0]}$ contains also the other \ion{He}{2} 
  resonance features. 
  Since the He lines and continua will be self-consistently computed in our 
  radiative transfer calculations, we need to exclude those contributions to 
  obtain the input values for our model calculations.  
   
  The \ion{He}{2}~256~\AA\ line has been observed to be at least 
  15 times weaker than the 304~\AA\ line in quiet regions 
  \citep{Mango-etal:78}. For an active region, 
  observations of the first four terms of the resonance series 
  indicate that the \ion{He}{2}~304~\AA\ is at least 40 times more intense 
  than the 256~\AA\ line, and about 25 times more intense than the sum of the 
  256, 243, and 237~\AA\ terms \citep{Thomas-Neupert:94}.  Evaluating the 
  contribution of the continuum at $\lambda<228$~\AA\ is considerably more 
  difficult \citep[see, for instance][]{Andretta-etal:03}.   
  In an optically thin recombination spectrum, the contribution of that 
  continuum could be as high as 30\% 
  of the total number of photons.  
  However, the existing observations of the resonance series (such as those 
  cited above) strongly suggest a formation of the \ion{He}{2} spectrum at 
  high optical depths.  Therefore, we may expect a considerably smaller 
  contribution of the continuum and of other lines, as already suggested by 
  \cite{Zirin:75} and \cite{Athay:88}.  Overall, we thus estimate that the 
  contribution of the \ion{He}{2} features below 256~\AA\ is less 
  than 10\% of that of the \ion{He}{2}~304~\AA\ line,  
  so that only the 
  latter line need be taken into account in the measured SEM irradiances. 
 
The contribution of the \ion{He}{2}~304~\AA\ line (plus the nearby  
\ion{Si}{11}~303~\AA\ line) to \SEM[1] irradiances has been estimated by 
\cite{Thompson-etal:02} at approximately 
50\%--60\% for observations acquired during 2001. 
Adopting an average value of 55\%, 
$\Irr{\mathrm{He}+\mathrm{Si}} = 
 1.4 \times 10^{10}$ photons cm$^{-2}$ s$^{-1}$   
 (the contribution of the \ion{Si}{11} 303~\AA\ line is 
 about 10\%). 
Subtracting this value from the \SEM[0] irradiances, we obtain  
the solar irradiance due to coronal lines only:   
$\Irr{\mathrm{corona}} \approx   
 3.64 \times 10^{10}$ photons cm$^{-2}$ s$^{-1}$. 
  The uncertainty in this value is mainly due to the uncertainty in the 
  absolute calibration of the \ion{He}{2}~304~\AA\ line relative to the other 
  lines in the CDS spectra 
  used by \cite{Thompson-etal:02}. 
  For CDS, the maximum uncertainty of the 
  2$^\textrm{nd}$ order calibration, relevant for the \ion{He}{2}~304~\AA\  
  line, can be conservatively estimated at $\approx 30\%$ \citep{Andretta-etal:03}.   
  Thus, the above value of 
  $\Irr{\mathrm{He}+\mathrm{Si}}$ has a maximum relative error of about 14\%, 
  which translates into a maximum error of $0.2 \times 10^{10}$ photons 
  cm$^{-2}$ s$^{-1}$ for $\Irr{\mathrm{corona}}$.  If we assimilate this 
  maximum error to a 3-$\sigma$ uncertainty, the standard deviation for 
  $\Irr{\mathrm{corona}}$ can be estimated to be $\approx 0.06 \times 10^{10}$ 
  photons cm$^{-2}$ s$^{-1}$.  We thus finally obtain: 
  $\Irr{\mathrm{corona}} =  
  (3.6\pm 0.1) \times 10^{10}$ photons cm$^{-2}$ s$^{-1}$ 
  (1-$\sigma$ error).  In the estimated uncertainty for this value, we can 
  safely include the errors introduced by neglecting the contributions of the 
  \ion{Si}{11}~303~\AA\ lines and of the others \ion{He}{2} features, 
  contributions which should be roughly of the same order, as mentioned above, 
  but of opposite sign. 

\subsection{Total number of ionizing photons at the solar surface}%
 \label{calc:photorad:total}

To obtain the contribution from the 
area around the DST slit to the total EUV solar irradiance   
we used the spatial 
distribution of EUV emission provided by EIT images. The details of the 
method are described in the Appendix and we illustrate here only the main 
assumptions and results. 
Basically, to obtain the mean value of  
$I^{\SEM[0]}$ and $I^{\SEM[1]}$ in any region of the EIT FOV,  
we scale the \SEM\ calibrated  
irradiances, \Irr{\SEM[0]} and \Irr{\SEM[1]},  
by the ratio of the mean counts in 
the region over the total counts in the full EIT FOV 
(Eq.~\ref{eq:fac_area_R}).  
 
This procedure assumes that the total counts over the detector in an EIT band 
(after basic CCD processing, including flat-fielding) 
are proportional to the irradiance, $\Irr{\mathrm{EIT}}$, in the same band. 
This is true if the contribution to $\Irr{\mathrm{EIT}}$ from the corona beyond 
1.2--1.4 $R_\sun$ (the limit of an EIT image) is negligible.  In fact, 
inspection of the EIT images used in our analysis reveals that only a small 
fraction of counts comes from areas above $1.1\: R_\sun$, and most of those 
counts are probably due to light scattered in the telescope. 
 
The strongest assumption in this procedure however, is that the images in the 
EIT wavebands are  good linear proxies of the intensity integrated within the 
\SEM\ wavebands.  In the case of the $\lambda<500$~\AA\ range (\SEM[0]), all 
EIT bands (around 171, 195, 284, and 304~\AA) include strong lines which 
contribute significantly to the total flux in that range. 
Thus, we may expect such an assumption to be quite reasonable.  The case of the 
\SEM[1] range is more straightforward: the EIT~304 waveband, 
although narrower, is centered at the same wavelength ($\approx$~300~\AA), and 
we may expect EIT~304 images to be good proxies of the spatial variation 
on the disk of the radiance in the \SEM[1] waveband. 
 
On the other hand, the assumption of linear correlation of EIT intensities with 
integrated intensities in the \SEM\ wavebands, is equivalent to assume that 
the spectral distribution of EUV photons in those wavebands does not change 
with the location on the Sun and that only the overall intensity of the lines  
changes.  This 
assumption is consistent with the calibration procedure of the \SEM\ data 
\citep{Judge-etal:98}, and  
is estimated to affect \SEM[0] measurements by about 10\%  
\citep{McMullin-etal:02}. 
 
Hence, we considered  a region of (70\arcsec)$^2$ centered at the position of the  
DST slit (solar rotation being taken into account), and we computed the  
conversion factors from \Irr{\SEM[]} to the mean value of $I^{\SEM}$ using 
two sets of EIT synoptic images taken before and after the DST and CDS 
observations (at 13:00 and 19:00 UT). Since we already subtracted the 
contribution of the \ion{He}{2}~304~\AA\ line to the \SEM[0] irradiance, we did 
not consider the EIT~304 images. In this way, we obtained an average conversion 
factor $c = 3.3 \pm 0.6\times 10^4\: \mathrm{sr}^{-1}$, where the  
uncertainty was obtained from the standard deviation of the distribution of  
counts within the region of interest. 
Therefore, the intensity of coronal lines, integrated in the $\lambda<500$~ 
\AA\ wavelength range and averaged over the area of the DST slit is  
$I^\mathrm{corona} = \Irr{\mathrm{corona}}\times c =  
(1.2\pm 0.2)\times 10^{15}$ photons cm$^{-2}$ s$^{-1}$ sr$^{-1}$.  
The uncertainty is  
obtained combining the uncertainties of $\Irr{\mathrm{corona}}$ and $c$. 
 
In order to validate this approach, we compare the intensity 
measured directly from CDS  in the considered area 
($I^{\mathrm{He}+\mathrm{Si}} = (1.1\pm 0.1)\times 10^{15}$) with the one 
obtained from the  \SEM[1] irradiance using the described procedure. 
The mean conversion factor from the two synoptic EIT~304 
images is 
$(5.8\pm 2.5)\times 10^4\; \mathrm{sr}^{-1}$ and therefore  
$I^{\mathrm{He}+\mathrm{Si}} $ 
$= \Irr{\mathrm{He}+\mathrm{Si}} \times c$ 
$= (0.8\pm 0.4)\times 10^{15}$ photons cm$^{-2}$ s$^{-1}$ sr$^{-1}$. 
The agreement between these estimates and the values measured with CDS is 
 well within the uncertainties.

\subsection{Angular and spectral distribution of photoionizing photons}%
 \label{calc:photorad:spectrum} 
 
The values computed in the previous section refer to intensities incident 
perpendicularly to the solar surface.  In our following  
calculations, we assume that the 
photoionizing incident intensities are constant with direction.  It is very 
difficult to evaluate this assumption, since 
the variation with angle of the photoionizing radiation depends on the spatial 
distribution of the coronal plasma around the region of interest.  For instance, 
had we assumed a vertically stratified, plane-parallel, optically thin corona, 
we would have obtained an intensity varying as $1/\mu$, where $\mu$ is the 
cosine of the angle with the normal.  Under such an assumption, 
the variation of the mean intensity, $J$, (and thus of the 
photoionization rate) with optical depth, $\tau$, is $J(\tau) \sim E_1(\tau)$, 
where $E_1(x)$ is the first exponential integral.  The constant intensity case 
we adopted gives instead a dependence $J(\tau) \sim E_2(\tau)$.  However, at large 
depths the two assumptions give the same result: $J(\tau) \sim 
\exp(-\tau)/\tau$. The differences are strong only in the shallower layers, 
with the ratio $E_2(\tau)/E_1(\tau)$ falling below 2 already 
for $\tau > 0.26$.

For the wavelength distribution of the ionizing photons, 
we adopt a synthetic spectrum obtained with the CHIANTI database 
\citep[version~4:][]{Young-etal:03} 
with the 
Differential Emission Measure (DEM) of an ``average'' active region, from 
\cite{Vernazza-Reeves:78}. 
In the calculations, we used 
standard elemental abundances, with some values scaled to mimic the FIP  
effect (further details in the file 
\texttt{sun\_hybrid\_ext.abund} in the CHIANTI package), and a value 
$P_\mathrm{e}/k = 3\times 10^{15}\; \mathrm{K}\; \mathrm{cm}^{-3}$, 
where $P_\mathrm{e}$ is the electron pressure and $k$ the Boltzmann 
constant. 
 
\begin{figure} 
  \centering 
  \resizebox{0.6\hsize}{!}{\includegraphics[angle=90]{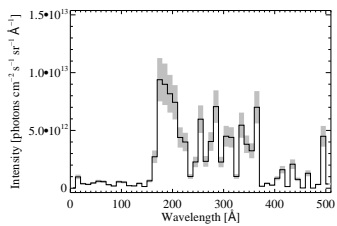}} 
  \caption{%
   Estimated spectral distributions of photons around the DST slit, derived as 
   described in Sec.~\ref{calc:photorad:spectrum}. The solid histogram shows 
   the spectral distribution derived from a CHIANTI synthetic spectrum; the 
   grey area indicates 1-$\sigma$ uncertainties.   
  } 
  \label{fig:EUVspec} 
\end{figure}

We need to scale the synthetic spectrum (excluding of course the 
\ion{He}{2} lines and continua) by a factor  
$1.3\pm 0.3$ 
to obtain the value of $I^\mathrm{corona}$ found in 
Sec.~\ref{calc:photorad:total}.  We note that such a linear scaling is 
equivalent to assuming that the shape of the DEM distribution of the plasma in 
our region of interest is the same as that of the distribution used to compute 
the synthetic spectrum. The resulting spectrum is shown in 
Fig.~\ref{fig:EUVspec}. 
 
\section{Atmospheric models}\label{calc:atmos} 
 
Our goal was the determination of 
semi-empirical models of the chromosphere and low transition region  
of the active region, providing a good match with the observed line profiles. 
The diagnostics we used for the chromospheric part of the models 
were the profiles of the Ca~{\sc ii} K, H$\alpha$, the Na~{\sc i} D lines, and  
the He~{\sc i} lines at 5876 \AA\ (D$_3$) and 10830 \AA, averaged over a  
region of $6\arcsec\times 3.4\arcsec$, which corresponds to the effective 
resolution of CDS. 
For the transition region we used the CDS profiles of the  
He~{\sc i} line at 584 \AA\ and of the He~{\sc ii} Ly-$\alpha$ line  
at 304 \AA. 
 
\subsection{General characteristics}\label{calc:atmos:charac} 
 
The modeling was done using the program PANDORA \citep{Avrett-Loeser:84}. 
Given 
a {\it T} vs. {\it z} distribution, we solved the non-LTE radiative transfer,  
and the statistical and hydrostatic equilibrium equations, and  
self-consistently computed non-LTE populations for 10 levels of H, 29 of  
He~{\sc i}, 15 of Fe~{\sc i}, 9 of C~{\sc i}, 8 of Si~{\sc i}, Ca~{\sc i} and  
Na~{\sc i}, 6 of Al~{\sc i} and He~{\sc ii}, and 7 of Mg~{\sc i}. In addition,  
we computed 6 levels of Mg~{\sc ii}, and 5 of Ca~{\sc ii}. 
 
More detail on the modeling, and on the different assumptions  
and their validity can be found in \cite{Falchi-Mauas:98}. 
Here, we just point out 
that the Ca II line-profiles are computed in Partial Redistribution and 
that line-blanketing is included considering Kurucz's opacities. 
The atomic models we used for H and Ca II are described in \cite{Mauas-etal:97} 
and \cite{Falchi-Mauas:98}. For He I, we used the 29 levels model described in 
\cite{Andretta-Jones:97} which, apart for the number of levels, differs very  
slightly from the model of  \cite{FAL:93} 
The different  
components in the lines were treated assuming that the sublevels are populated 
according to LTE, as explained in \cite{Mauas-etal:88}. 
For He II we used the 6 levels model of \cite{FAL:93}. 
 
The calculations include incident radiation from  
the coronal lines, which affects in particular the ionization balance in the  
H, He~{\sc i} and He~{\sc ii} continua. As explained in Sect.  
\ref{calc:photorad:spectrum}, this radiation was obtained scaling  
by a factor of 1.3 the distribution given by the CHIANTI database for an  
active region. As a starting point for the helium abundance, 
we used the standard photospheric value [He]=0.1. The  
microturbulence followed the semiempirical distribution given 
by \cite{FAL:91} 
as a function of total hydrogen density.

As studied by Fontenla et al. (2002), the inclusion of ambipolar diffusion 
of helium can have an important effect in the construction of a model, 
changing the energy balance and therefore the {\it T} vs. {\it z} structure. 
However, it must be remarked that their models of the low transition region  
are theoretical. We have made some trial runs including ambipolar diffusion 
but  keeping our atmospheric structure unchanged, and found that the emitted  
profiles remain unchanged. 
 
\subsection{Model construction}\label{calc:atmos:standard} 
 
Given the different sensitivity of various spectral lines  
to modifications in different parts of 
the model, we can describe its construction as a step by step process.   
 
The source function for H$\alpha$ is 
sensitive to the structure of the chromosphere from the temperature 
minimum region to the base of the transition region, even if the line 
center is formed higher up in the atmosphere. The Ca~{\sc ii} K line, 
in particular its wings and the K$_1$ minimum, are formed at the 
temperature minimum and in the low and mid-chromosphere. Therefore, we 
fixed the deeper part of the chromospheric structure using these two 
profiles, and cross-checked it with the profile of the Na D lines, which are 
much less sensitive to these atmospheric layers.  None of these 
lines depend on the radiation field. 
 
The profiles of the He~{\sc i} lines at 5876 \AA\  
(D$_3$) and 10830 \AA\ are formed in two distinct regions. Most of the  
radiation in both lines is originated in the photosphere, which in the quiet  
Sun results in a weak absorption line at 10830 \AA\, and no noticeable  
line at 5876 \AA. However, in an active region there is also a 
chromospheric contribution which depends mainly on the coronal EUV  
incident radiation, but 
also on the thermal structure of the high chromosphere/low transition region,  
between $1.\times 10^4$ and $2.5\times 10^4$ K. We used these lines, therefore, to fix  
this region as a second step in the construction of the model. 
 
Finally, both ultraviolet lines, the He~{\sc i} line at 584 \AA\ and the  
He~{\sc ii} line at 304 \AA, depend  
on the structure 
of the low and mid-transition region, from $3.\times 10^4$ to $5.\times0^4$ K for the 584 \AA\  
line, and up to $1.\times 10^5$ K for the 304 \AA\ line. As mentioned in the 
Introduction, their dependence on the 
radiation field is not yet assessed, and should therefore be investigated. 
Hence, we used these lines in a final step as diagnostics of the structure 
of the transition region. 
 
\begin{figure} 
   \centering 
   \includegraphics[width=7cm]{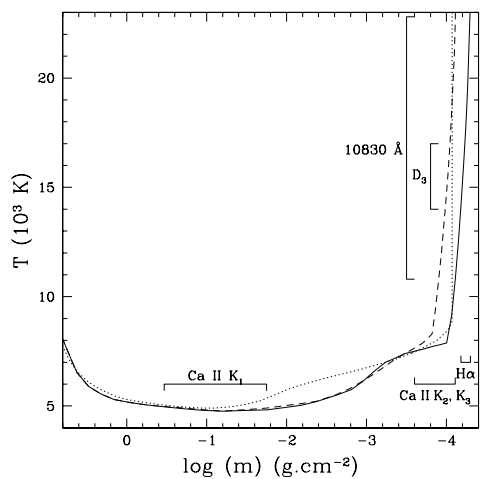} 
   \caption{Temperature vs column mass distribution of the computed 
 atmospheric models. The models shown were obtained with the 
 microturbulent velocity of \cite{FAL:91}  (dashed line), and with a 
 value of 5 km/s in the region of formation of the lines (solid line, see 
 text). 
 The plage model P \citep{FAL:93}  is displayed for reference 
 (dotted line).  Also shown are the heights of formation of the 
 different lines we used to build the models. The He~{\sc i} 584 line 
 is formed at temperatures between 
 $2.4\times 10^4$ and $3.4\times 10^4$ K, and the He~{\sc ii} 304 line is formed between 
 $9.5\times 10^4$ and $1.\times 10^5$ K.} 
   \label{mod_vt} 
\end{figure}

In Fig.~\ref{mod_vt} we show the distribution of the temperature as a function  
of column mass for the model  
that gives the best match to the observations (dashed line).  
Also shown are the regions of  
formation of the different lines we used as diagnostics for the model  
construction. 
 
\begin{figure} 
   \centering 
   \includegraphics[width=7cm]{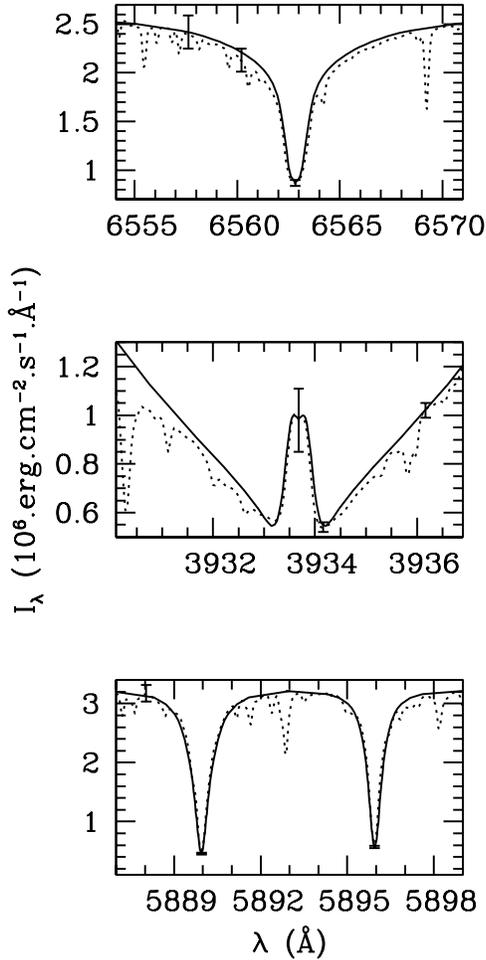} 
   \caption{Observed (dashed line) and computed (solid line) profiles for 
   H$\alpha$ and the Ca II K and Na D lines. The bars indicate the r.m.s.  
   of the averaged profiles (see text) and point out the 
   variability between pixels.} 
\label{prof_hcn} 
\end{figure}

In Fig.~\ref{prof_hcn} we show the computed and observed line profiles for  
H$\alpha$, the Ca~{\sc ii} K and Na D lines, which do not depend on the  
radiation field. All computed profiles were convolved with the instrumental 
response. The bars in the Figure indicate the r.m.s. 
of the profiles averaged over the area observed by CDS. It can be seen that the  
agreement found is very good, well whithin the r.m.s. 
 
\begin{figure} 
   \centering 
   \includegraphics[width=7cm]{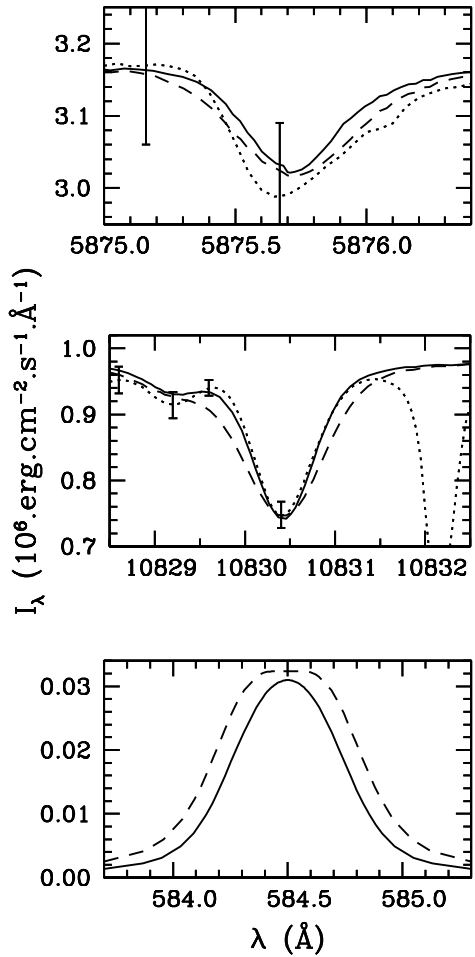} 
   \caption{Dotted line: Observed profiles for three helium lines.  
   Dashed: computed profiles (convolved with instrumental response) for the model 
   adopting the microturbulence of \cite{FAL:91}.  Full line: 
   same as dashed, for our standard model where microturbulence has been 
   reduced (see text).  In the upper and middle panel, bars indicate  
   the r.m.s. of the averaged 
   profiles and point out the variability between pixels. 
   For the shallow D$_3$ line (maximum central depth $\approx$ 5\%) the 
   variability is very large and the bars are clipped at the limits of 
   the plot.} 
   \label{prof_hel} 
\end{figure}

In Fig.~\ref{prof_hel} we compare the observed helium lines with  
the profiles obtained from this model, 
again convolved with the instrumental response.  
We want to stress that for  
the 584 \AA\ the instrumental effects are 
rather heavy, washing out completely 
the strong central self-absorption of the computed profile (see following 
Section and Fig. \ref{prof_584_dist}). The \ion{He}{2}~304 \AA\ line profile (not 
shown) is very similar to the 584 \AA \ line.  
For all the helium lines, although the central 
intensities agree with the observations within the variability between pixels,  
the computed profiles result too broad.  
Furthermore, the observed profile of the 10830 \AA\ line distinctively shows 
the weak blue component of the triplet, which in the  
computed profile appears washed out. 
 
The main parameter that contributes to the broadening of these lines 
is the microturbulent velocity (v$_t$). As described above, in this 
model we adopted the turbulence distribution given  
by \cite{FAL:91}. 
To obtain a better match with the observed He lines we then  
tried a different microturbulence 
distribution, changing v$_t$ from 10 to 5 km s$^{-1}$ in the region with  
temperature ranging from $1.\times 10^4$ to 2$1.\times 10^4$ K.  
Since the microturbulent velocity is included in the hydrostatic equilibrium  
equations, changing it implies changing the densities, and therefore the  
atmospheric structure. Hence, the whole procedure was repeated, until a 
satisfactory match with the observations was found, and a new model was built.  
This model is also shown in Fig.~\ref{mod_vt} as a solid line, and  
the corresponding profiles for  
the He lines are shown in Fig.~\ref{prof_hel}. The new profiles for  
H$\alpha$, the Ca~{\sc ii} K and Na D lines (not shown) are very similar to the 
ones obtained with the former model. It can be seen that the match to He 
profiles is now strongly improved. We have therefore adopted this model as our 
standard model for the observed active region.  
 
Finally, we checked {\it a posteriori} that the value of $P_e/k=3\times  
10^{15}$ K cm$^{-3}$, assumed to compute the spectral distribution of the coronal 
radiation with the CHIANTI database, agrees with that found with our
models. The most external point of the model for 
which we have an observational constraint, has a temperature 
T$=1.1 \times 10^5$ K  and an electron density
N$_e=2.3\times 10^{10}$ cm$^{-3}$, 
resulting in a very good agreement with $P_e/k=2.5\times 10^{15}$ K cm$^{-3}$ .

\subsection{Effect of the EUV radiation}\label{calc:atmos:EUVspectrum} 
 
\def\lam{} 
 
\begin{figure} 
   \centering 
   \includegraphics[width=7cm]{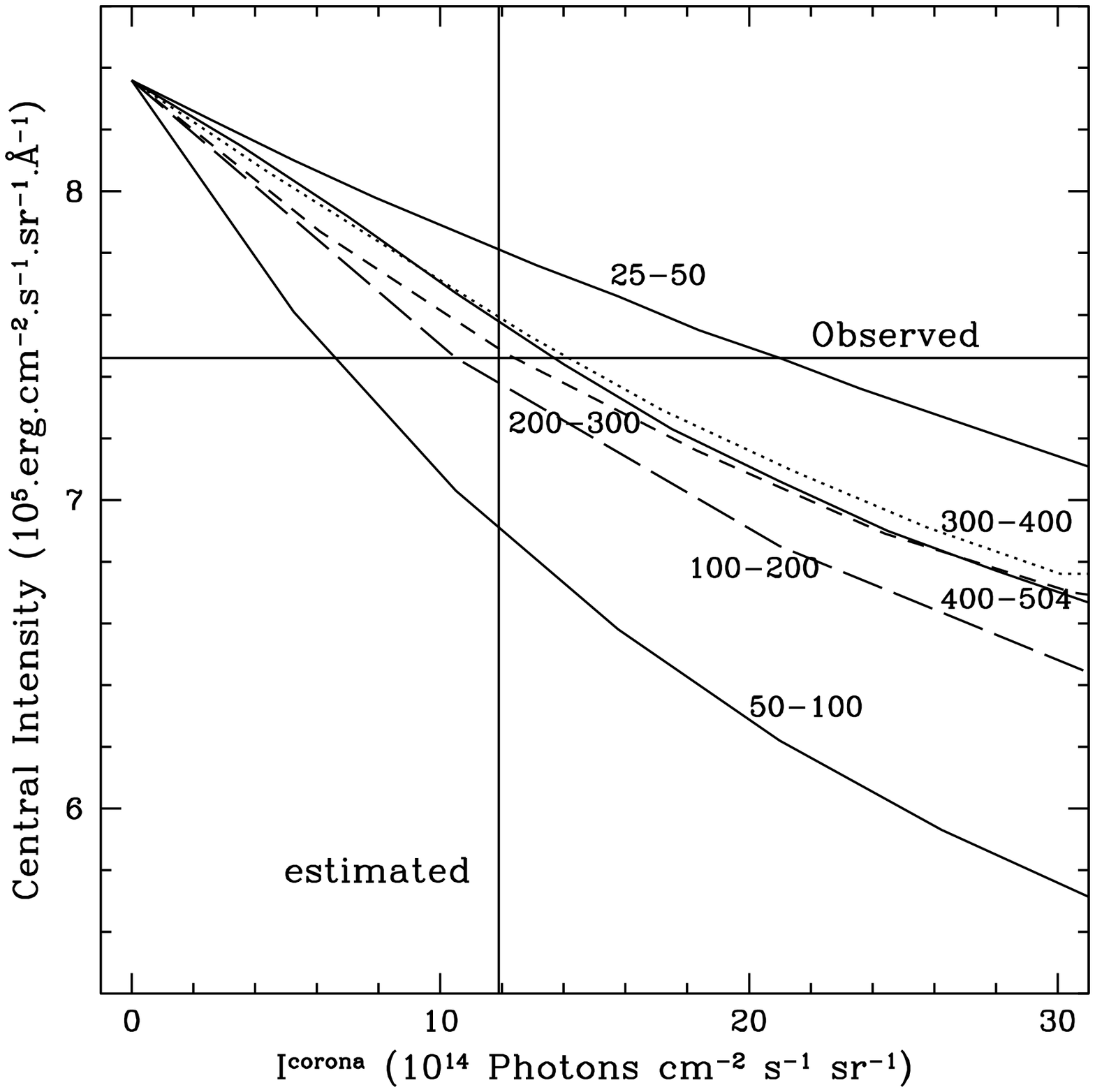} 
   \caption{Central intensity for the \lam 10830 \AA\ line as a function of  
     the intensity $I^\mathrm{corona}$ for different step distributions of the 
     field.} 
   \label{int_vs_phot} 
\end{figure}

To assess the influence of the coronal radiation on the profiles of the helium 
lines, we performed new calculations for different intensities of the 
radiation field, 
and for different spectral distributions.  
Since the most important change in 
the profiles with the radiation field is in the central depth of the D$_3$  
and of the \lam\ 10830 \AA\ line, in Fig.~\ref{int_vs_phot}  we show the intensity  
in the center of the 10830 line as a function of the intensity  
$I^\mathrm{corona}$.
 
Each curve in the figure represents a different spectral 
distribution. In each case, we considered a step function, which is non-zero 
only in a limited spectral range, and zero outside it. In this way, we tested 
distributions with all the photons localized between 25 and 50 \AA, between 50  
and 100 \AA, 100 and 200 \AA, 200 and 300 \AA, 300 and 400 \AA\ and between 400 
and 504 \AA. 
 
Several important conclusions can be drawn from this Figure.  
  First, photons at short wavelength ($\lambda < 50$~\AA) have a small 
  influence on the line profiles. Second, for the same value of 
  $I^\mathrm{corona}$, photons between 50 and 100 \AA\ are the most effective 
  in increasing the line central intensity. Third, the details of the spectral 
  distribution between 100 and 504 \AA\ are of little relevance for the 
  resulting central intensity. 
   
  Such effects, anticipated by \cite{Andretta-etal:03} and 
  reprised in Sect. \ref{calc:photorad}, can be understood in terms of the 
  dependence on wavelength of the fraction of EUV photons effectively 
  capable of photoionizing helium atoms and ions.  The first effect is 
  essentially due to the fact that short-wavelength photons are almost 
  entirely absorbed by metals, via inner-shell photoionizations.  At longer 
  wavelengths, the relevant parameter is the ratio of helium and 
  hydrogen photoionization cross sections.  The fact that the ratio between 
  50 and 100~\AA\ is about 3--5 times the same ratio at 500~\AA, is consistent 
  with the second conclusion drawn from Fig.~\ref{int_vs_phot}.  Likewise, the 
  third result can be understood by noting that the ratio between the 
  \ion{He}{1} and hydrogen cross-sections changes in a relatively slow fashion 
  at long wavelengths. 
   
  In light of these results, we note that, in principle, any given central 
  intensity of the 10830 line  
  could be obtained with different values of $I^\mathrm{corona}$, 
  but only if assuming very different spectral distributions. For example, the 
  observed central intensity can be obtained with 8 $\times 10^{14}$ photons 
  cm$^{-2}$ s${^-1}$ sr$^{-1}$ between 50 and 100 \AA, or with almost twice as 
  many photons between 300 and 504 \AA.  However, for a ``stationary'' active region 
  such as in our case, the hypothesis of a dominant contribution at shorter 
  wavelengths is rather unplausible, so that our assumption of a 
  synthetic spectrum obtained with the CHIANTI database (Sect. 
  \ref{calc:photorad:spectrum}) seems fully justified. 

\begin{figure} 
   \centering 
   \includegraphics[width=7cm]{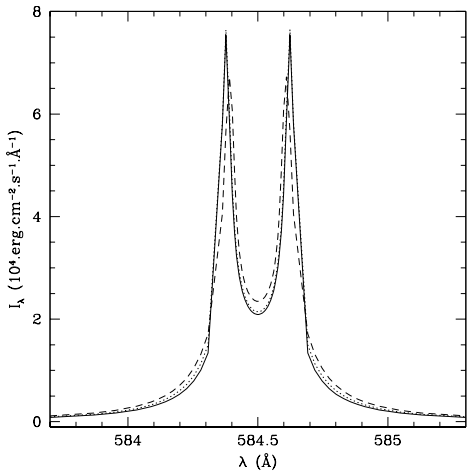} 
   \caption{584 \AA\ profile computed with no 
   incident coronal radiation (full line),  compared with the 
   profiles computed for  
   $I^\mathrm{corona} =$  
    1 $\times 10^{15}$ photons cm$^{-2}$ s$^{-1}$ sr$^{-1}$  
   but with different spectral distribution. 
   Dashed line: all photons between 400 and 504 \AA. Dotted line: all 
   photons between 50 and 100 \AA.} 
   \label{prof_584_dist} 
\end{figure}

In Fig.~\ref{prof_584_dist} we show the effect of the coronal 
radiation on the 584 \AA\ profile. We compare the profiles for the 584 \AA\  
obtained with no incident coronal radiation with those obtained with  
$I^\mathrm{corona} =$  
    1 $\times 10^{15}$ photons cm$^{-2}$ s${^-1}$ sr$^{-1}$ 
with two different spectral distribution. We see that the profiles are 
very similar and that the maximum difference 
(in the peak intensity) is about 10\%. The profile variation depends  
more on the 
spectral distribution than on the intensity of coronal radiation.  
As expected, when  
all photons are between 50 and 100 \AA\ and  penetrate deeper in the 
atmosphere, the line does not change appreciably.  
A small change occurs instead  when  
all photons are between 400 and 504 \AA. 
We obtained a 
similar result for the He~{\sc ii} 304 \AA\ line. 
 
Our analysis hence shows that, similarly to the quiescent case,  
also for active regions the coronal radiation has a much stronger effect 
on the chromospheric He lines than on the TR He lines. 
 
\subsection{Helium Abundance}\label{calc:atmos:abun} 
 
In the previous Sections, we have shown how a semi-empirical model 
employing the standard value of [He]=0.1 can reproduce very well the 
observational data. On the other hand, the actual value of [He] in different 
parts of the solar atmosphere is a matter of discussion (see Introduction), so 
it might be of interest to investigate how different values of [He] affect our 
computations. 
 
\begin{figure} 
   \centering 
   \includegraphics[width=7cm]{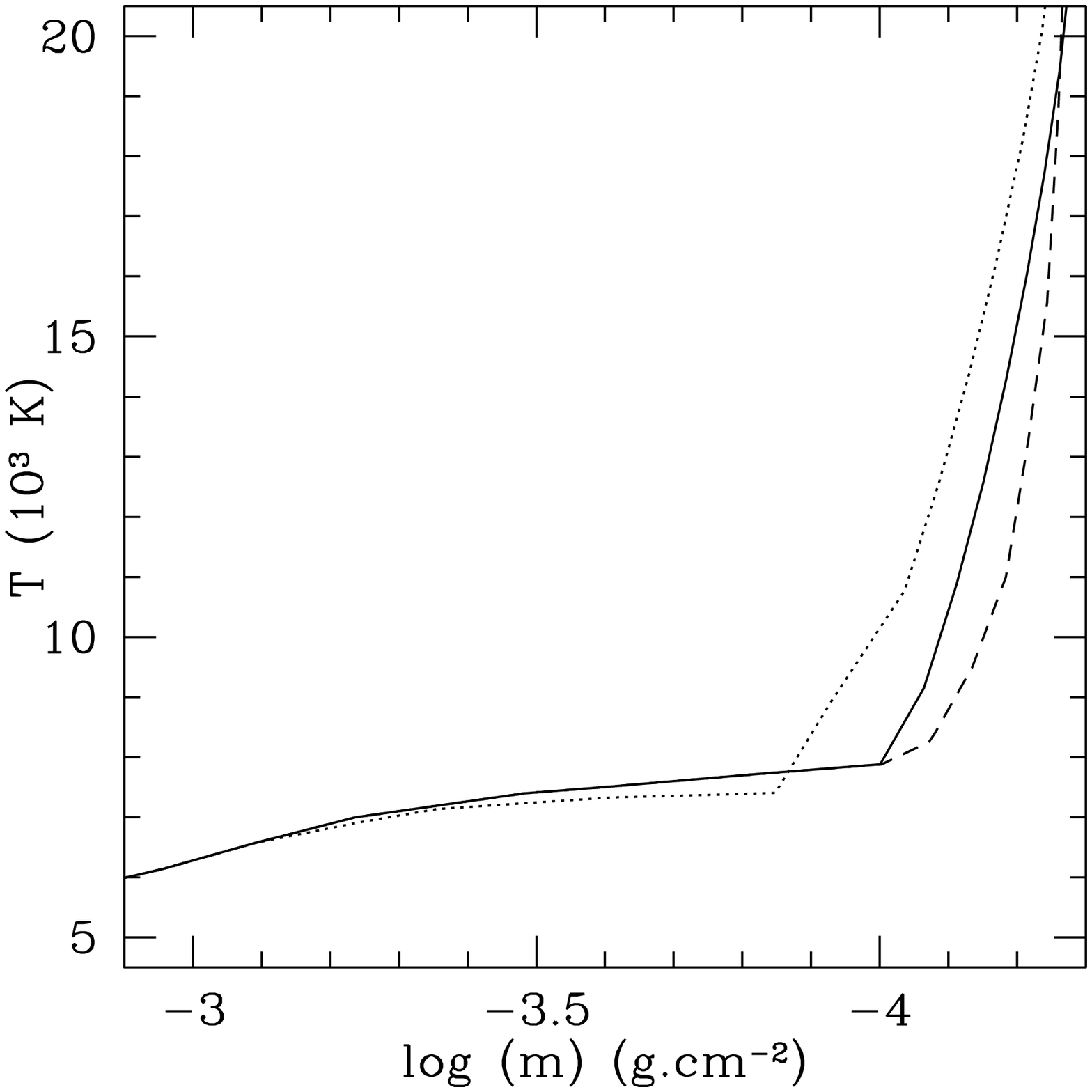} 
   \caption{Temperature vs column mass distribution of the atmospheric 
   models obtained for different values of the He abundance, in the region  
   where they differ. Solid line: [He]=0.1, our standard model. Dotted line: 
   [He]=0.07. Dashed line: [He]=0.15. } 
   \label{mod_abund} 
\end{figure} 
 
Changing the abundance modifies the total density 
for a given height, and therefore affects the hydrostatic equilibrium. Hence, 
modifying [He] in the whole solar atmosphere would result in a completely  
different atmospheric structure. In particular, the resulting photosphere 
could not be constrained by our observations. Furthermore, the region  
between the chromosphere and transition region has been indicated as a good 
candidate for processes that might be responsible for strong variations of 
[He] \citep{Lie-Svendsen-etal:03}.  
For these reasons, we   
changed [He] only from the point where T = $1.\times 10^4$ K outwards, 
where the He lines are formed, and recomputed the models until we 
found a satisfactory match with the observations.  
We built two different models, assuming values of the abundance of [He]=0.07 and  
[He]=0.15.  
 
\begin{figure} 
   \centering 
   \includegraphics[width=7cm]{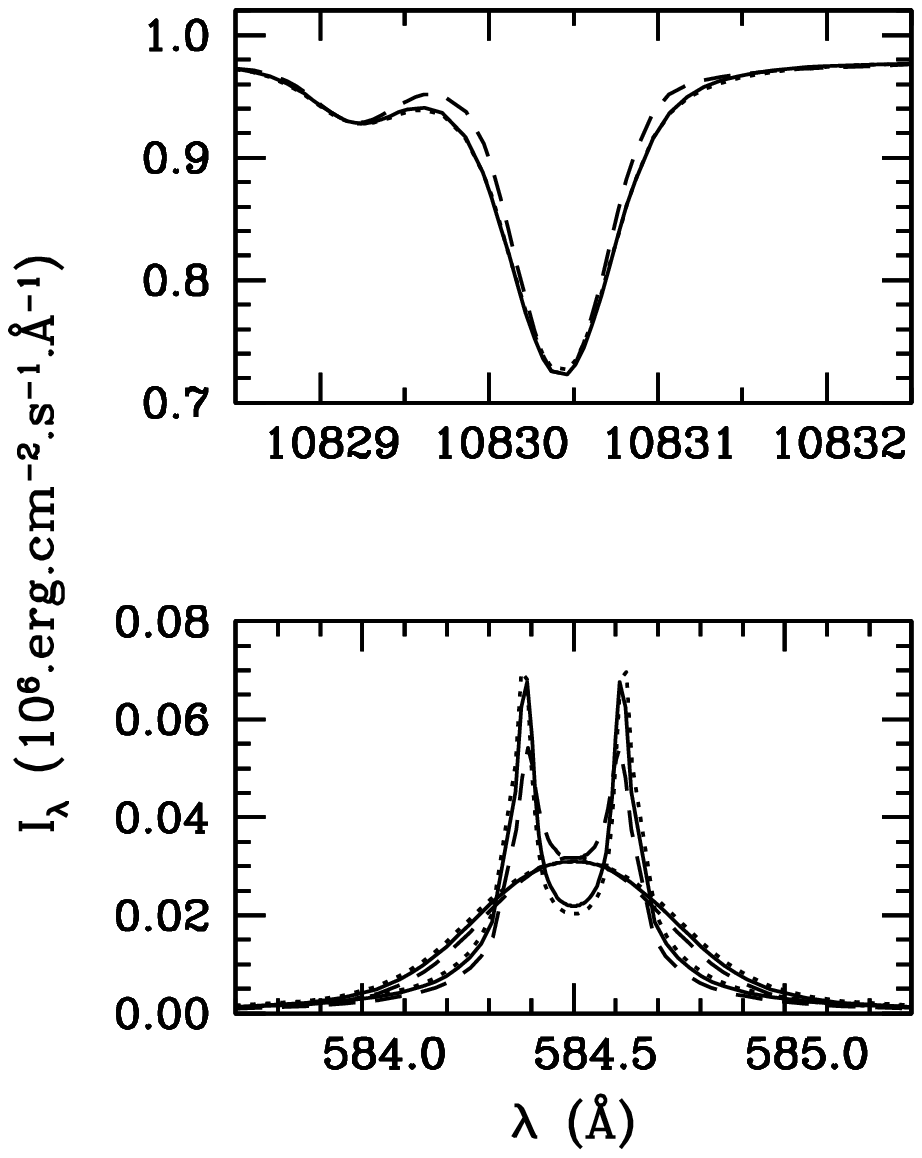} 
   \caption{Profile for two helium lines, for different values of the helium 
   abundance. Solid line: [He]=0.1, our standard model. Dotted line: 
   [He]=0.07. Dashed line: [He]=0.15.  
   For the 584 \AA\ also the profiles before the convolution with the 
   instrumental response are shown.} 
   \label{prof_abund} 
\end{figure}

In Fig.~\ref{mod_abund} we show these models, together with the one obtained  
for the standard abundance [He]=0.1, in the region above log(m)=-3, where  
they differ. In Fig.~\ref{prof_abund} we compare the 10830 and 584 \AA~ 
profiles obtained with these three models. In general,  
the differences in the profiles are very small, although a larger value of abundance 
corresponds to narrower profiles in both lines. 
This is due to the fact that to compensate for a smaller abundance the region 
of formation of the lines needs to be extended, to obtain enough helium 
atoms. This larger region between $1.\times 10^4$ and $2.5\times 10^4$ K 
can be noted in 
Fig.~\ref{mod_abund}, and results in a broader line.  
Although the ratio between the peaks and the central absorption of the 584 \AA\ changes 
by a factor of two with the extreme abundance values, the 
convolution with the broad CDS response destroys any difference. 
Only observations of EUV lines at much higher spectral resolution might then 
help in discriminate between different abundance values.

\section{Discussion and conclusions}\label{disc} 
 
We obtained  
a large set of simultaneous and cospatial observations of an active 
region, including 
 the chromospheric lines Ca~{\sc ii} K, H$\alpha$ and Na~{\sc i} D as 
 well as the \ion{He}{1} lines at 5876 (D$_3$), 10830 and 584 \AA\ and the 
 \ion{He}{2} line at 304 \AA. 
The EUV radiation in the range $\lambda<500$~\AA\ and in the range 
$260<\lambda<340$~\AA\  has  also been measured at the same time. 
Using the program PANDORA we were able to build  
semiempirical atmospheric models to match this set of observables. 
The He lines and continua are self-consistently computed in the radiative 
transfer calculations. In particular we were interested in the role of the 
photoionization-recombination process in the formation of the He lines and  
in the effect of possible variations of [He] in an active region. 
 
One of the main problems for the calculation of the He line profiles 
was to estimate the total number of ionizing photons impinging on the 
target active region, and their spectral distribution. The adopted 
procedure combines the observations from CELIAS/\SEM\ and EIT 
instruments for our region with the synthetic spectrum obtained with  
the CHIANTI database for an ``average'' active region and a 
$P_\mathrm{e}/k = 3\times 10^{15}\; \mathrm{K}\; 
\mathrm{cm}^{-3}$. We needed to 
scale the synthetic spectrum by a factor  
$1.3\pm 0.3$ 
to obtain the estimated value of $I^\mathrm{corona}$  
(Sec.~\ref{calc:photorad:total}).  
 
We have two independent checks to validate this approach. First, we compared 
the \ion{He}{2} 304 intensity, measured  
directly from CDS in the considered area, with the one obtained from 
the  \SEM[1] irradiance in the range $260<\lambda<340$~\AA\, and the  
agreement resulted well within the 
uncertainties. Second, {\it a posteriori},  
we found that the electron pressure for the most 
external point of the model for which we have an observational 
constraint, agrees with the one assumed to compute the spectral 
distribution with the CHIANTI database. The agreement guarantees that the 
obtained model is self-consistent with the EUV coronal intensity 
used. 
 
A semiempirical model with a standard value for [He] and a modified 
distribution of microturbulence v$_t$  
(reduced in the region of formation of the He lines) 
provides a good agreement with all the 
observations. However, we must stress that the broad CDS response function 
heavily conceals the details of the EUV line profiles, thus 
limiting their diagnostic relevance.  
 
For this model, defined as our standard model, we study the influence 
of the coronal radiation on the computed helium lines.   
We find that, similarly to the quiescent case, also in an active region the incident 
coronal radiation has a limited effect on the UV He lines,  
while it results fundamental for the D$_3$ and 10830 lines.  
For these latter two lines, the photons of wavelengths 
below 100 \AA\ result very effective in increasing the line depth,  
confirming that for the calculation of the He line 
profiles it is crucial to correctly estimate the total number of ionizing photons  
and their spectral distribution. While 
the assumption of an average distribution might not be critical in 
the case of a ``stationary'' active region,  
this could be not true in the case of a flare, where the coronal EUV 
radiation is probably harder. 
 
Finally, we tested how the helium abundance influenced our computed 
profiles, changing   
the value of [He] only in the region where 
the temperatures are larger than $1.\times 10^4$ K. 
We built two more models for [He]=0.15 and [He]=0.07  
and found that also with these models we could  
match the observations as well as with the standard model. The differences 
in the computed lines are mostly evident for 584 and 304 \AA\, whose ratio  
between the peaks and the central absorption changes 
by a factor of two with the extreme [He] values. However, given the 
coarse spectral resolution of CDS, they are not 
appreciable in our observations that, hence, do not provide enough 
constraints to choose between [He] values. 
Observations of the 584 and 304 \AA\ lines with a
spectral element smaller than 0.025 AA/pix around 584 \AA \ might help in
this issue. This resolution
is currently achieved only by the SUMER spectrograph aboard SOHO. However,
SUMER has been restricted to off limb observations since year 2000.
Moreover it does not observe the He II Lyman alpha around 304 \AA. Thus we
have to wait for a next generation EUV spectrographs aboard one of the
fortcoming missions such as Solar Orbiter. 
Furthermore,  
since the changes of the atmospheric structure with [He] 
are larger in the TR, it would be useful to constrain  
this part of the atmosphere with other observables, independent on the  
He lines. In particular we suggest that simultaneous observations 
of EUV lines of C~{\sc ii}, C~{\sc iii}, Si~{\sc ii} and Si~{\sc iii} 
might be  very important to define the structure of the TR and hence help to 
discriminate between [He] values. 
 
 
\acknowledgments 
 
We thank  D.\ McMullin, W.\ Thompson, G.\ Del Zanna
for their help and
useful advices about obtaining, analyzing, and interpreting the SOHO
data used in this paper.
We would like to thank the CDS and NSO teams for their unvaluable support in 
performing these observations. 
 
SOHO is a project of international cooperation between NASA and ESA. 
 
CHIANTI is a collaborative project involving NRL (USA), RAL (UK), and the 
Universities of Florence (Italy) and Cambridge (UK).

 
\appendix 
 
\section{Derivation of intensities from irradiance measurements}%
 \label{app:conv} 
 
If the Sun were a featurless disk, with no limb brightening nor off-limb 
emission, the disk-integrated flux (irradiance), $\Phi$, measured at a distance 
$D$ would be easily related to the specific intensity, $I$, at the solar 
surface: 
\begin{displaymath} 
  \Phi = I \times \pi R_\sun^2/D^2 = I \times \Omega_\sun ,  
\end{displaymath} 
where $R_\sun$ and $\Omega_\sun$ are, respectively, the solar radius, and the 
angular dimension of the solar disk (in steradians). 
 
 
In the case of a spatially resolved, full-disk solar image where it is possible 
to identify $N$ discrete structures, each of area $S_i$ ($i=1\ldots N$), and 
emitting a surface intensity $I_i$ along the line of sight,  
the observed irradiance is: 
\begin{displaymath} 
  \Phi = \sum_i I_i \times \frac{S_i}{D^2} = \sum_i I_i \times \Omega_i, 
\end{displaymath} 
where $\Omega_i = S_i/D^2$ is the angular size of the $i$-th 
structure. 
 
>From the above relation, the contribution to the irradiance from the individual 
structure $i$ is therefore: 
\begin{displaymath} 
  \Phi_i = \left(\frac{I_i \times \Omega_i}{\sum_j I_j \times \Omega_j}\right) 
           \times \Phi .  
\end{displaymath} 
Hence,  
the surface specific intensity in that structure can be written as  
\begin{equation}\label{eq:int_area} 
  I_i = \frac{\Phi_i}{\Omega_i} = c_i \times \Phi\ , 
\end{equation} 
where the conversion factor $c_i$ (in units of an inverse solid angle) is 
\begin{equation}\label{eq:fac_area} 
  c_i = \frac{I_i}{\sum_j I_j \times \Omega_j}\ . 
\end{equation} 
 
The relations~\ref{eq:int_area} and \ref{eq:fac_area} are useful if the total 
irradiance is available in absolute units (e.~g.\ photons cm$^{-2}$ s$^{-1}$), 
while a solar image in the same bandpass provides spatially resolved,  
uncalibrated (e.~g.\ counts pixel$^{-1}$ s$^{-1}$), intensities.  It is also 
worth noting that the structures in Equations~\ref{eq:int_area} and 
\ref{eq:fac_area} can be arbitrary: the intensities $I_i$ can in fact be the 
intensities of a single pixel, or of a set of pixels. 
 
In the limit of continuous intensity distributions, it is easy to generalize 
those relations to obtain the mean intensity in an arbitrary region ${\cal R}$ as 
\begin{displaymath} 
  \langle I\rangle_{\cal R} \equiv \frac{1}{\Omega_{\cal R}}  
                                   \int_{{\cal R}}\!I\:\mathrm{d}\Omega\ . 
\end{displaymath} 
Hence, the generalization of relations~(\ref{eq:int_area}) and 
(\ref{eq:fac_area}) is trivial: 
\begin{equation}\label{eq:int_area_R} 
  \langle I\rangle_{\cal R}^\mathrm{calib.} =  
   \Phi^\mathrm{calib.}\times\langle c\rangle_{\cal R}, 
\end{equation} 
where the mean conversion factor is 
\begin{equation}\label{eq:fac_area_R} 
  \langle c\rangle_{\cal R} =  
  \frac{1}{\Omega_{\cal R}}\ , 
  \frac{\int_{{\cal R}}\!\mathrm{d}\Omega\:I^\mathrm{counts}} 
       {\int\!\mathrm{d}\Omega\:I^\mathrm{counts}}\ .  
\end{equation}

 
\bibliographystyle{apj} 
\bibliography{helium_regatt_vin.bib} 
\end{document}